\documentclass[prl,aps,twocolumn,showpacs,superscriptaddress,groupedaddress]{revtex4-1}
\usepackage{amsmath}
\usepackage{amssymb}
\usepackage{textgreek}
\usepackage[pdftex]{graphicx}
\usepackage{subfigure}
\DeclareMathOperator{\var}{var}
\DeclareMathOperator{\E}{E}
\usepackage[some]{background}

\SetBgScale{1}
\SetBgContents{\parbox{20cm}{%
  \Huge Preprint of article published in Biological Cybernetics\\[18.5cm]\rotatebox{180}{\Huge }}}
\SetBgColor{black}
\SetBgAngle{270}
\SetBgOpacity{0.2}

\begin{document}
\title{On the Rate Coding Response of Peripheral Sensory Neurons}
\author{Willy Wong}
\email{willy.wong@utoronto.ca}
\affiliation{Department of Electrical and Computer Engineering, University of Toronto, Toronto M5S3G4}
\affiliation{Institute of Biomaterials and Biomedical Engineering, University of Toronto, Toronto M5S3G9}
\date{\today}

\begin{abstract}
The rate coding response of a single peripheral sensory neuron in the asymptotic, near-equilibrium limit can be derived using information theory, asymptotic Bayesian statistics and a theory of complex systems.  Almost no biological knowledge is required.  The theoretical expression shows good agreement with spike-frequency adaptation data across different sensory modalities and animal species.  The approach permits the discovery of a new neurophysiological equation and shares similarities with statistical physics.   \end{abstract}

\maketitle
 \BgThispage
\section{Introduction}
Sensory transduction is the process whereby sensory stimuli are converted to neural responses.  The sensory system is the gateway to the brain, and the processing of information its principal task.  The mathematical relationship between information and the peripheral sensory response is a topic of current interest.  

This paper attempts to show that the rate coding response of a peripheral sensory neuron in the so-called asymptotic, near-equilibrium (or near steady-state) limit can be characterized by a single equation of information with no detailed knowledge of the underlying physiology.  The basis of this approach is that the sensory system performs a measurement process involving the estimation of a sensory signal.  The entropy of this estimate is then attributed to the response of the neuron.  This is all that is required to characterize sensory processing at its most elementary level.  

The theory presented here concerns the problem of \textit{rate coding}.  Increasing stimulus magnitudes typically induce higher rates of response.  Moreover, the response of a neuron to a steady signal drops monotonically over time, a process known as adaptation.   The proposed theory works well with response of single isolated units in the sensory periphery when probed with slowly varying stimuli of sufficient intensity and is broadly applicable to different organisms and modalities.  However, the success of the theory can depend on several factors which are discussed later.

This paper is a continuation of a series of publications detailing an information or entropy approach to sensory processing \cite{norwich1977information,norwich1993information}.  From this theory, aspects of sensory science can be unified using a Boltzmann or Shannon measure of uncertainty together with a few auxiliary assumptions.  The approach was later extended to neurophysiology \cite{norwich1995universal,wong1997physics}.  The current paper goes further to demonstrate that the asymptotic, near-equilibrium sensory response can be derived using ideas from information theory, asymptotic Bayesian estimation and complexity theory. There have been other studies which have looked at the coding of neural information (e.g. \cite{benda2003universal,drew2006models,aviel2006spiking,famulare2010feature}).  But generally they do not tackle the problem of responses in primary afferent neurons, and can be more complex than the current approach.

\section{Derivation of main equations}
\subsection{Overview}
The derivation of the equations governing the sensory response can be summarized as follows: The receptor samples the sensory signal to estimate the mean.  The uncertainty in the mean can be quantified with information theory resulting in an expression connecting measurement uncertainty with the variance of the signal and the number of samples drawn.  A key assumption is the association of firing rate with measurement uncertainty: that is, the greater the uncertainty, the greater the firing rate.  Finally, using a theory of complex systems together with an assumption of how quickly sampling takes place, an equation emerges connecting firing rate with stimulus intensity and duration.  This equation can then be compared to experimental data like adaptation responses measured from peripheral neurons.  

\subsection{Detailed derivation}
Let $\theta$ denote the parameter estimated by the sensory system.  In the case of intensity coding, $\theta$ refers to the magnitude of sensory stimulation.  The sensory receptor draws repeated, independent samples $X$ from an unknown distribution, i.e. $X_1, X_2,...,X_m \sim p(x|\theta)$.  Given the prior distribution $\pi_0(\theta)$ (representing the uncertainty in $\theta$ before any measurements), after $m$ samples the posterior distribution takes the form
\begin{equation}
\pi(\theta)=p(\theta | X_1,...,X_m )\propto p(X_1,...,X_m|\theta) \pi_0(\theta) \label{bayes}
\end{equation}
This is an expression of Bayes' theorem.  In the limit of large $m$, and under most conditions observed in nature, the posterior distribution is asymptotically normally distributed with mean parameter equal to the maximum likelihood value $\hat{\theta}$ and variance proportional to $\var(X)/m$,
\begin{equation}
\pi(\theta)\overset{d}{\longrightarrow} \mathcal{N}\left(\hat{\theta},\var(X)/m\right) \label{posterior}
\end{equation}
where $\var(X)$ is the variance of the sensory signal.  The form of the asymptotic distribution is independent of the choice of the prior.  This result is sometimes referred to as the Bernstein-von Mises theorem \cite{van2000asymptotic} and is discussed in greater detail below.

Stimulus samples are processed with limited resolution.  We assume the error to be normally distributed with zero mean and variance $R$.  The entropy is calculated from the mutual information obtained from the posterior and the error distributions.  This is achieved by taking the convolution of the posterior and error distributions, calculating its entropy, and then subtracting from it the entropy of the error distribution alone to give
\begin{equation}
H=\dfrac{1}{2}\log\left(1+\frac{\var(X)}{m R}\right) \label{hfunction}
\end{equation}
This is then the expression governing uncertainty in the sample mean. Equation (\ref{hfunction}) is identical in form to the Shannon-Hartley law for an additive white Gaussian noise channel with signal-to-noise ratio equal to $\var(X)/mR$ \cite{cover2012elements}.  

Equation (\ref{hfunction}) was first derived in the context of sensory processing over forty years ago \cite{norwich1977information}.  The original derivation made use of the central limit theorem to derive the asymptotic form of the distribution of uncertainty in $\theta$.  In this paper, we use instead a Bayesian approach which makes clear the role of the prior distribution.  The derivation of the posterior distribution in (\ref{posterior}) requires a number of steps.  Following \cite{van2000asymptotic}, the asymptotic form of the posterior distribution for $m \rightarrow \infty$ can be shown to have mean equal to $\hat{\theta}$ and variance equal to the reciprocal of the Fisher information of $\theta$.  In the case where $X$ belongs to the one-parameter exponential family (which includes most of the well-known random variables observed in nature) and $\theta$ is a natural parameter of the family, there exists an efficient estimator of $\theta$ which achieves the Cram\'{e}r-Rao lower bound \cite{doksum2007mathematical}.  In this case, the reciprocal Fisher information equals $\var{\left(X\right)}/m$.  For the sensory problem considered here, $\theta$ is the signal magnitude, and the \textit{sample mean} obtained from $X_1, X_2,...,X_m$ is an efficient estimator of $\theta$.  Thus, implicit in this approach is the idea that the sensory receptor averages to estimate intensity.

By itself, (\ref{hfunction}) has already many of the characteristics required to describe mathematically the process of intensity coding.  Given a constant sensory signal, an increase in the number of samples or measurements $m$ results in a monotonic reduction of uncertainty $H$.  Recall that during adaptation the sensory response to a steady input also falls monotonically.  This suggests that entropy $H$ can be related to the sensory response through the equation
\begin{equation}
F=kH \label{ffunction}
\end{equation}
where $F$ is the firing rate or spike response of a neuron and $k$ is a positive constant with units of spikes per second.  The fall in neural response during adaptation can be interpreted as a gain in certainty in the sensory signal.  When the uncertainty vanishes, there is no response.  The association of firing rate with measurement uncertainty also permits the testing of theory with experimental data.  For extensive discussion and the origins of this equation see \cite{norwich1977information,norwich1983perceive,norwich1993information}.  While other choices of a monotonic relationship between firing rate and uncertainty are possible, (\ref{ffunction}) is the simplest and provides an interesting analogy between sensory processing and statistical physics. 

The use of information in this approach differs fundamentally from how information is used typically within neuroscience (e.g. \cite{kostal2018coordinate}).  In this paper, information is not a measure of capacity, but a calculation of uncertainty from which a physiologically measurable response (i.e. firing rate) is determined.  This is not unlike how entropy is calculated in physics with a mathematical model of molecular uncertainty using Boltzmann's $H$-function, which is then related back to the thermodynamic system via $S=-k_B H$.

When (\ref{ffunction}) is combined with (\ref{hfunction}), we see that the rate coding response of a sensory unit must increase monotonically with signal variability.  Is this prediction supported by experimental observation?  For example, the phenomenon of brightness enhancement (aka the Br\"{u}cke-Bartley effect, e.g. \cite{marks1974sensory}) shows that the apparent brightness of a flickering light can change with the frequency of flicker when time-average luminance is kept constant.  However flickering contributes to temporal variations in the signal resulting in the enhancement in apparent brightness.  Other experiments involving the stabilization of an image on the retina show that prolonged exposure to a fixed image leads to the fading of the visual percept, e.g. \cite{ratliff1950involuntary}.  In each case, we observe that the sensory response is coupled to variations in the signal.  There are also other approaches which take sensation to be coupled to variation or changes in the signal, e.g. \cite{laming1986sensory,itti2009bayesian}.  

However, neither of the above experiments probe the exact relationship between variance and \textit{firing rate}.  Instead a new experimental test can be proposed to test this assumption directly.  Light exhibits very different statistical behaviour depending on whether it is in the classical or quantum limit.  Photon bunching is the phenomenon whereby the statistics of the photon count deviates from a Poisson distribution (e.g. \cite{paul1982photon}).  If a photoreceptor is stimulated with such a signal, the resulting neural response can be recorded to test the dependency of firing rate on variance with mean held constant.

Yet it is clear even from the classic studies of Adrian and Zotterman \cite{adrian1926impulsesb} that the neural response encodes the \textit{mean} of the signal, i.e. the intensity.  An increase in mean generally results in an increase in neural response. As such, we expect the dependency of $F$ to be on $\E(X)$ and not $\var(X)$.  How can this discrepancy be resolved?  Some recent work has shown that many complex systems exhibit a power-law relationship between mean and variance.  The fluctuation scaling law was first discovered in ecology through animal population studies and is known also as Taylor's law \cite{taylor1961aggregation}.  A compelling explanation for the fluctuation scaling law was recently proposed \cite{kendal2011taylor}.  The family of probability distributions known as the Tweedie distributions exhibits a power law relationship between the mean and variance.  A convergence theorem has been established suggesting a reason for the ubiquity of the power law in complex systems \cite{jorgensen1997theory}.  

Let us assume first the applicability of the fluctuation scaling law to sensory signal statistics.  Its implications will be discussed later.  Introducing $\var(X)=\epsilon \mu^p$, where $\epsilon$ and $p$ are positive constants and $\mu=\E(X)$ and defining a new constant $\beta=\epsilon/R$, we obtain
\begin{equation}
H=\dfrac{1}{2}\log\left(1+\frac{\beta \mu^p}{m}\right) \label{interm}
\end{equation}
The response is now a monotonic increasing function of the mean.  See \cite{norwich1977information} for the original derivation of this equation.  

The signal mean consists of both external and internal sources.  The external source is the sensory signal itself and any other external environmental signals. Internal sources may include self-generated signals, as well as stochastic neural noise (e.g. \cite{kuhn2004neuronal,schwalger2010noisy}).  We model the mean as a sum of the two components $\mu=I+\delta I$ where $I$ is the total magnitude of external sources and $\delta I$ the sum of internal sources or noise.  $\delta I$ will generally be small relative to the external input for almost the entire range of $I$.

Next we consider the role of time in the sensory response.  Sample size increases with the number of measurements taken.  Hence $m$ is a function of time and $dm/dt$ refers to the \textit{sampling rate}.  It is reasonable to assume that the sample size does not increase indefinitely: when the number of samples attains the optimal value, the sample size remains constant.  Sampling is thus a function of the difference between the current sample size $m$ and the optimal value $m_{eq}$.  That is,
\begin{equation}
\frac{dm}{dt}=g(m-m_{eq}) \label{sample}
\end{equation}
where $g$ is some function with the condition $g(0)=0$.  Near equilibrium, we take a Taylor expansion around $m = m_{eq}$ to obtain
\begin{align}
\frac{dm}{dt}&\simeq g(0)+\dot{g}(0) (m-m_{eq}) \\
&=-a (m-m_{eq})
\end{align}
Since the number of samples $m$ must be less than $m_{eq}$ and $dm/dt \ge 0$, $a=-\dot{g}(0)$ is a positive time constant.  Solutions of $m$ are used to calculate $H$ from (\ref{interm}) given a choice of $m_{eq}$.

One final step is required before the derivation is complete.  The determination of the optimal sample size $m_{eq}$ will depend on the precise condition for optimality.  In Appendix \ref{section:optimal}, it is shown that if the estimation error is constrained then the optimal sample size $m_{eq}$ must grow as a function of stimulus intensity in the form of a power function.  See equation (\ref{meq}).

Summarizing, we have
\begin{align}
&F=kH \label{gut1}\\
&H=\dfrac{1}{2}\log\left(1+\dfrac{\beta \left(I+\delta I\right)^p}{m}\right) \label{gut2}\\
&\dfrac{dm}{dt}=-a (m-m_{eq}) \label{gut3}\\
&m_{eq}=(I+\delta I)^{p/2} \label{gut4}
\end{align}
A list of the key assumptions used in the derivation can be found in Appendix \ref{section:assumptions}.

From the approximations above, it is expected that the equations work best at the near-equilibrium, near steady-state limit where $m \simeq m_{eq}$, and at intensities large enough such that (\ref{posterior}) is satisfied (i.e. asymptotic normality).  Nevertheless, we shall see that these equations give a good description of the neural response to most time-varying sensory inputs for intensities up to the physiological saturation levels.  

\section{Discussion}
The derivation of (\ref{gut2}) requires the use of a Tweedie distribution with $\var(X)=\epsilon \E(X)^p$.  Tweedie distributions belong to the exponential family and exist for all real values of $p$ except $0<p<1$ \cite{jorgensen1997theory}.  This turns out to have important consequences for the growth of the neural function.  Compression is an essential property of sensory neurons since sensory signals can range over several orders of magnitude while the dynamic range of a peripheral neuron is far more limited.

In the asymptotic limit of large sample size where $m=m_{eq}$, the following can be derived from (\ref{gut1}-\ref{gut4}) by taking a Taylor series expansion of the logarithm
\begin{equation}
F=\frac{k \beta}{2}  (I+\delta I)^{p/2} \label{feq}
\end{equation}
A compressive response involves a power exponent less than one.  Since $p$ itself is positive, and no such Tweedie model exists for $0<p<1$, this implies that the only possible range of exponents is $p \in [1,2)$.  Such Tweedie models are known as \textit{compound Poisson-gamma models}  \cite{jorgensen1997theory}.  A compound Poisson-gamma model can be generated via a sum of gamma-distributed random variables, with the number of summed terms itself Poisson distributed.

Fluctuation scaling would thus imply that the interaction between signal and receptive field is well-characterized by a Poisson-gamma model when the response is compressed relative to the range of input.  In the olfactory system, for example, odourant molecules bind with receptor sites on the cilia in the epithelial layer \cite{pifferi2009signal}.  At steady-state, the number of binding events per unit interval of time is likely Poisson distributed.  The number of receptor sites activated is a cluster and cluster sizes are often modelled by gamma distributions.  It would appear that the Poisson-gamma model provides not only a reasonable model for olfaction, but for other modalities as well.  For sensory modalities where the range of stimuli is more limited (e.g. mechanoreception or stretch sensing), the neural response may not necessarily be compressive.  When $p>2$, this would imply that $X$ has a distribution belonging to the family of positive stable distributions \cite{jorgensen1997theory}. 

\section{Predictive scope of the equations}
The equations governing sensory entropy can be solved for different inputs and experimental configurations using the method of first-order transient analysis.  The more challenging step is to find experimental paradigms which allow for the robust determination of the five unknown parameters found in (\ref{gut1})-(\ref{gut4}).  These parameter values are not predetermined and are specific to receptor type, as well as to individual units.  To avoid overfitting, we make use of the idea that multiple experiments conducted on the same unit should obey the same set of parameters.  This is a stringent test of the theory as it greatly reduces the number of degrees of freedom allowed to the equations.  

\subsection{Time-independent inputs}
\begin{figure}
\centering
\includegraphics[width=0.4\textwidth]{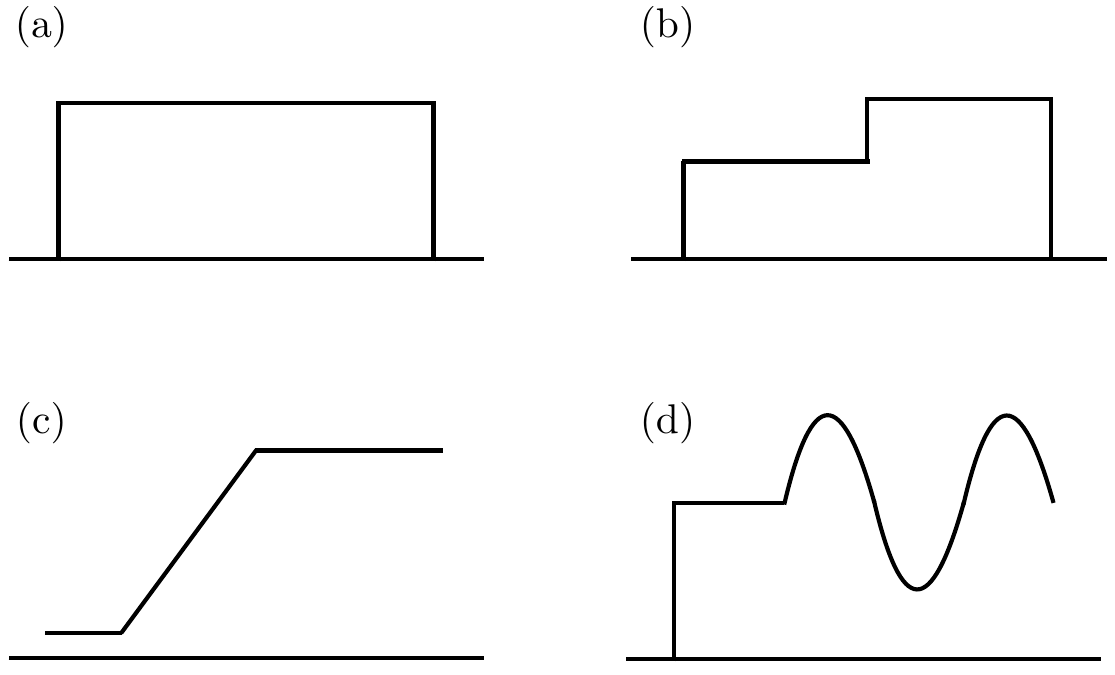}
\caption{A schematic illustration of sensory inputs commonly used to probe the response of sensory units. In all figures, the ordinate shows firing rate and the abscissa time.  (a) A step input is used to measure adaptation.  (b) Double-step function.  (c) A ramp-and-hold stimulus. (d) Sinusoidally modulated intensity superimposed on a constant background.}
\label{fig:fig1}
\end{figure}

We begin with examples involving piece-wise constant inputs.  First consider the solution for a step input illustrated in Figure \ref{fig:fig1}a.  We divide the solution into three distinctive regions: region 1 ($t<0$) where the stimulus is off, region 2 ($0 \le t< t_0$) where stimulus is turned on, and region 3 ($t \ge t_0$) where the stimulus is again off.

Next the relevant response is solved.  The general solution of the first order ordinary differential equation in (\ref{gut3}) is given by
\begin{equation}
m(t) = m(t') e^{-a\left(t-t' \right)}+a e^{-at} \int\displaylimits_{t'}^{t} e^{a\tau} m_{eq} d\tau \label{m(t)}
\end{equation}
when the initial condition is evaluated at $t'$. For constant $I$, $m_{eq}$ is constant and thus (\ref{m(t)}) can be simplified to
\begin{equation}
m(t) = m(t') e^{-a\left(t-t' \right)}+m_{eq} \left[1-e^{-a\left(t-t' \right)} \right]
\end{equation}
Assuming that the neuron is equilibrated (i.e. fully adapted) prior to $t<0$, the sample size $m(t)$ can now be solved in all three regions to give
\begin{align}
&m_\text{1}=m_{eq1}\label{begin}\\
&m_\text{2}=m_\text{2}(0)e^{-at}+m_{eq2} (1-e^{-at}) \label{middle}\\
&m_\text{3}=m_\text{3}(t_0)e^{-a(t-t_0)}+m_{eq3} \left[1-e^{-a(t-t_0)}\right]\label{end}
\end{align}
where $m_{eq1}=m_{eq3}=\delta I^{p/2}$ and $m_{eq2}=(I+\delta I)^{p/2}$. The values for $m_{eq}$ were obtained by substituting for intensity in (\ref{gut4}).  Finally, continuity ensures that $m_\text{2}(0)=m_\text{1}(0)$ and $m_\text{3}(t_0)=m_\text{2}(t_0)$.  Substituting $m$ and $I$ into (\ref{gut1}) and (\ref{gut2}) gives the response of the neuron in all three regions.  Other inputs (e.g. a double-step input in Figure \ref{fig:fig1}b) can be solved similarly.

We will now compare the theory with experimental data.  All curve-fits were conducted in MATLAB R2019b (Mathworks) using the function {\tt nlinfit} unless otherwise noted.

\subsubsection{Auditory adaptation and driven activity}
\begin{figure}
\centering
\includegraphics[width=0.48\textwidth]{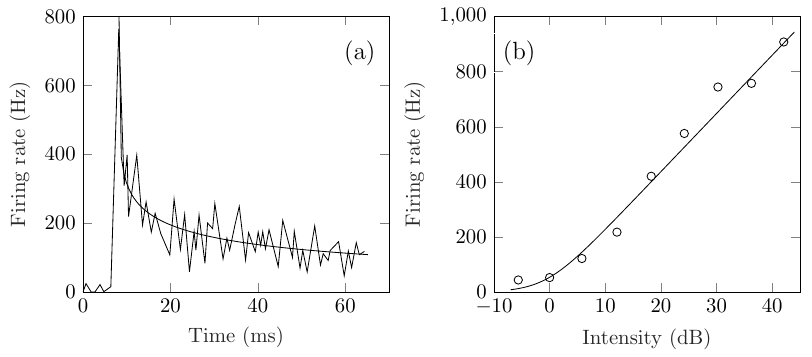}
\caption{Firing rate response recorded from the auditory fibre of a Mongolian gerbil \cite{smith1988encoding}.  Data in both figures recorded from the same fibre.  Smooth curves show the predictions of (\ref{adapt}) using a common set of parameters for both figures.  (a) Firing rate measured as a function of sound duration for a 39 dB tone (jagged line).  (b) Peak firing rate measured as a function of sound intensity in decibels (open circles).}
\label{fig:fig2}
\end{figure}

Figure \ref{fig:fig2} shows data from two sources: an adaptation experiment (constant $I$, duration $t$ is varied) and an intensity-rate experiment (constant $t$, $I$ is varied) \cite{smith1988encoding}.  Data was recorded from the same auditory fibre of an anesthetized Mongolian gerbil (unit MB-52-11 from Figures 4 and 6).  In the adaptation experiment, the number of spikes counted in a 960 $\mu$s interval was converted to a firing rate and observed as a function of time.  An averaged firing rate was obtained over 91 trials.  Figure \ref{fig:fig2}a (jagged line) shows the response to a 39 dB SPL tone presented at the characteristic frequency of the unit (2.44 kHz).  In the intensity-rate experiment, the maximal firing rate during a one millisecond interval was recorded as a function of different sound intensities.  Figure \ref{fig:fig2}b shows the intensity-rate response curve (open circles).  After 40 dB, the response saturates and is not shown in Figure \ref{fig:fig2}b.  

The expression for $F$ used to fit the data was derived from the sample size in region 2, i.e. (\ref{middle}), and is given by
\begin{equation}
F=\frac{1}{2}k\log\left[  1+\frac{\beta\left(  I+\delta I\right)  ^{p}}{\delta I^{p/2} e^{-at}+\left(  I+\delta I\right)  ^{p/2}\left(  1-e^{-at}\right)
}\right] \label{adapt}
\end{equation}
Since both experiments were conducted on the same auditory unit, a common set of five parameters was used  ($k=1.3\times10^2 $, $\beta=2.2\times 10^{-3}$, $p=2.8$, $\delta I=1.0\times 10^{-4}$, and $a=5.3\times 10^{-3}$ Hz).  Stimulus intensity in dB was calculated from rms pressure relative to 20 $\mu$Pa.  For the intensity-rate curve, $t$ was set equal to zero.  Figure \ref{fig:fig2} shows good compatibility between theory and data.  

\subsubsection{Auditory double-step input}
\begin{figure}
\centering
\includegraphics[width=0.48\textwidth]{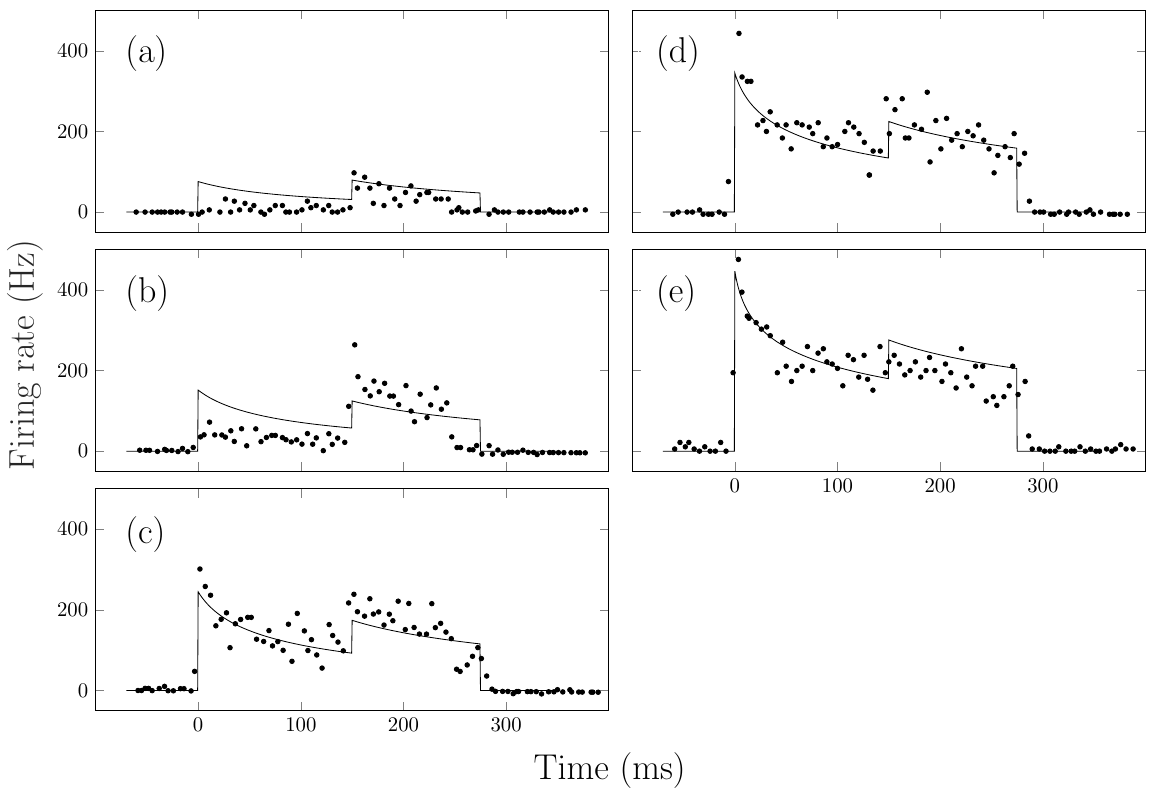}
\caption{Auditory neural responses measured from a double-step input in the guinea pig ear \cite{smith1975short}.  Initial pedestal with intensity $-4$, 2, 8, 14 or 20 dB (panels a-e respectively), followed by a second pedestal 6 dB higher.  Firing rates indicated by filled circles.  Smooth curves show the predictions of the equations solved for a double-step input using a common set of four parameters for all five graphs.}
\label{fig:fig3}
\end{figure}

In \cite{smith1975short}, the response was measured to a series of double-step inputs from the auditory nerve of guinea pigs (data from Unit GP-6-2, see Figure 3 from original paper).  A schematic illustration of the input is shown in Figure \ref{fig:fig1}b.  The initial response was elicited with a sound of intensity $-4$, 2, 8, 14, or 20 dB SPL followed by a 6 dB increase in the second pedestal.  The low spontaneous activity and slow rate of adaptation allows us to take a Taylor series approximation for both $\delta I \ll I$ and $t \ll 1/a$.  The solid line in Figure \ref{fig:fig3} shows the predictions with four adjustable parameters ($k=1.6\times 10^{2}$, $\beta/a=1.5\times 10^{2}$, $\delta I^{p/2}/a=4.2\times 10^{1}$, $p=1.8$).  A weighting function of $F^{1/4}$ was introduced in the fitting procedure.  The match is not perfect although the theoretical curves largely capture the behaviour observed physiologically (filled circles).

\subsubsection{Peak versus steady-state response}
In the same study, peak response of the adaptation curve was compared to steady-state response over a range of intensities (data of Unit GP-17-4 shown in Figure 1 of \cite{smith1975short}).   These results can be used to test a key component of the theory: that the optimal sample size grows with intensity following (\ref{gut4}).  In an attempt to reduce the number of parameters in (\ref{adapt}), we approximate the equation by taking the large intensity limit.  This is achieved by ignoring the `$1+$' term in (\ref{adapt}) and taking internal noise to be small (i.e. $I \gg \delta I$) to obtain
\begin{equation}
F = \frac{1}{2}k\log\left[  \frac{\beta I  ^{p}}{\delta I^{p/2} e^{-at}+ I ^{p/2}\left(  1-e^{-at}\right)
}\right] \label{approxadapt}
\end{equation}
The peak response PR is calculated by setting $t=0$ and the steady-state response SS with $t \rightarrow\infty$:
\begin{align}
&\text{PR}= \frac{1}{2}k\log\left(\beta I  ^{p}\right)-\frac{1}{2}k\log \left(\delta I^{p/2}
\right) \label{peak}& \\
&\text{SS}= \frac{1}{2}k\log\left( \beta I  ^{p/2}\right)& \label{ss}
\end{align}
These two quantities, when plotted against the logarithm of intensity (or decibel), will yield two lines with slope differing by a factor of two.  In total, three fitting parameters were required (two intercepts and one slope).  Figure \ref{fig:fig3} shows the predictions together with the experimental results from \cite{smith1975short}.

Most peripheral sensory units will exhibit a peak response to steady-state relationship predicted by (\ref{peak}-\ref{ss}) although there are exceptions, e.g. \cite{benda2005spike}.  See also Section \ref{section:new} which shows the same relationship but in a different form.

\begin{figure}
\centering
\includegraphics[width=0.28\textwidth]{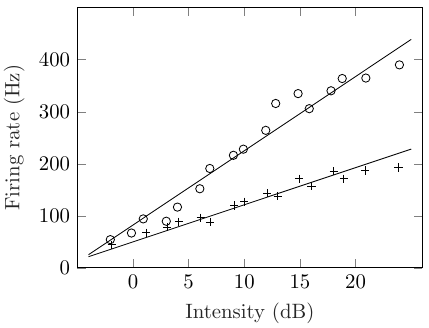}
\caption{Responses recorded from the guinea pig ear showing peak activity (open circles) and steady-state activity (plus signs) from adaptation curves \cite{smith1975short}.  Solid lines show the predictions of the theory.  The two lines are expected to differ by a factor of two in slope.}
\label{fig:fig4}
\end{figure}

\subsubsection{Multiple olfactory adaptation}
The adaptation response in the sugar receptors of blowflies was measured for three different concentrations (0.01, 0.1 and 1 M, from Figure 4 of \cite{dethier1984relations}).  The experiment was conducted in the region where the adaptation had not yet reached steady-state.  The concentrations are assumed to be sufficiently high such that $\delta I$ can be ignored.  As such, we used a simpler version of (\ref{adapt}) by taking $\delta I=0$ and carrying out a first order Taylor series expansion for $t \ll 1/a$ in the denominator to obtain
\begin{equation}
F = \frac{1}{2}k\log\left[  1+\frac{\beta' I^{p/2}}{t}\right] \label{classicadapt}
\end{equation}
where $\beta'=\beta/a$.

\begin{figure}
\centering
\includegraphics[width=0.28\textwidth]{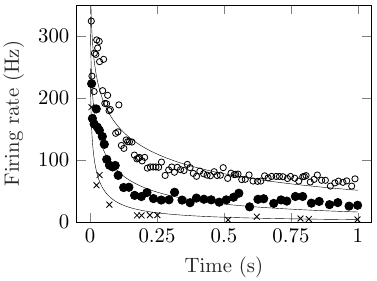}
\caption{Adaptation responses recorded in the sugar receptor of a blowfly for three concentrations (1.0 M open circles, 0.1 M filled circles, 0.01 M crosses) \cite{dethier1984relations}. The responses were recorded from the same unit.  Smooth curves indicate a simultaneous curve-fit with (\ref{classicadapt}) using the same three parameters for all three data sets. Figure and result adapted from \cite{norwich1991informational}}
\label{fig:fig5}
\end{figure}

This equation holds special significance as it is the original form of the equation governing sensory response based on the entropy approach. First published in 1977, it was the first attempt to use entropy to connect together various empirical sensory laws and appeared in a number of publications (e.g. \cite{norwich1977information,norwich1993information}).  For example this equation encompasses two of the famous empirical equations used to describe sensory data.  For weak intensities $\beta' I^{p/2}/t \ll 1$, a power law emerges, whereas for larger intensities $\beta' I^{p/2}/t \gg 1$, the Weber-Fechner logarithmic law is obtained.  The fit shown in Figure \ref{fig:fig5} was originally published in 1991 \cite{norwich1991informational}.  In total, three curves were fitted using three unknown parameters ($k=1.1\times 10^2$, $\beta'=1.5\times 10^3$ and $p=1.3$).  

\subsection{Time-varying inputs}
Hitherto, we have considered responses to inputs that are piece-wise constant.   In general, analytical solutions for time-varying inputs are not possible due to the exponent in (\ref{gut4}).  However, numerical solutions can be easily obtained by either solving the differential equation in (\ref{gut3}) using an Euler method or through numerical integration. 

\begin{figure}
\centering
\includegraphics[width=0.28\textwidth]{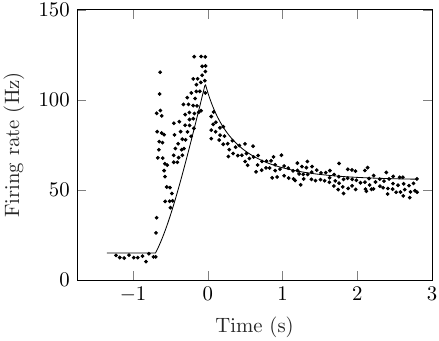}
\caption{Neural response to the ramp-and-hold lengthening of a cat muscle fibre \cite{schafer1994regularity}.  The theoretical prediction requires a numerical solution to the differential equation in (\ref{gut3}).  In total, 4 parameters were used to fit three experimental regions: the initial steady-state region, ramp and subsequent adaptation response.}
\label{fig:fig6}
\end{figure}

\subsubsection{Muscle ramp-and-hold}
In this example, the response in a cat to muscle spindle lengthening was recorded to a ramp input (see Figure 1 of \cite{schafer1994regularity}).  The fibre was elongated linearly and then held fixed at its final length.  A schematic representation of the input is shown in Figure \ref{fig:fig1}c.  The stimulus in this case is a time-varying function.  The solution for the sample size $m$ was obtained by solving (\ref{gut3}) numerically.  In an attempt to reduce the number of parameters, the small intensity limit of (\ref{gut2}) was adopted by taking the linear approximation $\log(1+x) \simeq x$.  The parameters $k$ and $\beta$ combine to become a single parameter.  In total, 4 parameters were used for 3 experimental regions ($k\beta=0.20$, $\delta I=1.8\times 10^{-4}$, $p=4.5$ and $a=1.4$ Hz).  A weight function of $1/F^2$ was used with the fit.

\subsubsection{Sinusoidal variation and adaptation response in mechnoreception}
Recordings were taken from the slit sense organ of a hunting spider \cite{bohnenberger1981matched}.  Two different adaptation responses were recorded together with the response to a sinusoidal input from the same type of mechanoreceptor unit.  (Both experiments were conducted on slit 2 of the lyriform organ although there is no indication of whether the recordings were made from the same receptor unit or not.  See Figures 4 and 6 from original paper.)  The mechanoreceptor responded only to the positive half of the sinusoidal input which is typical for mechanoreception.  Adaptation responses were calculated using (\ref{adapt}) with inputs $0.0975^\circ$ and  $0.395^\circ$, while the sinusoidal input was evaluated through a numerical solution of (\ref{gut3}) with a 1 Hz sinusoidal input with peak value $0.25^\circ$.  Five parameters were used for three different experiments ($k=31$, $\beta=5.8$, $p=6.1$, $\delta I= 6.6\times 10^{-6}$ and $a=1.8\times10^{-3}$ Hz).

\begin{figure}
\centering
\includegraphics[width=0.48\textwidth]{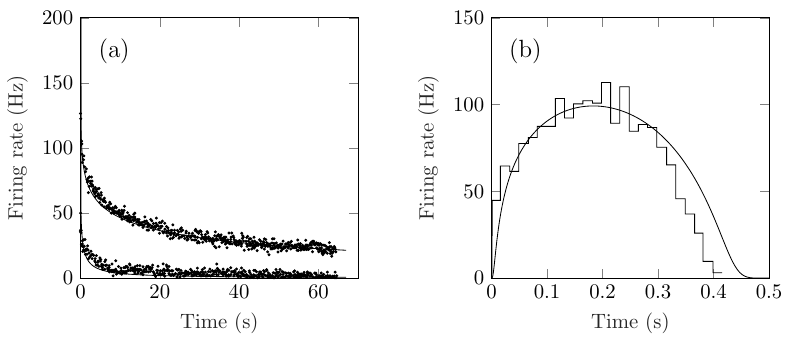}
\caption{Neural responses recorded from possibly the same mechanoreceptor unit of a hunting spider \cite{bohnenberger1981matched}.  (a) Two adaptation responses to differing intensities (upper curve $0.395^\circ$, lower curve $0.0975^\circ$) together with predictions of (\ref{adapt}).  (b) Response to a 1 Hz sinusoidal input with peak deflection $0.25^\circ$ (jagged line) with numerical solution of (\ref{gut1}-\ref{gut4}) (smooth line).  A total of five adjustable parameters were used for all three experiments.}
\label{fig:fig7}
\end{figure}

\subsubsection{Square pulse versus sinusoidal responses in muscle spindle}
The response of spindle afferents to repeated square pulse stimulation was compared to sinusoidal stimulation in the soleus muscle of cats (Figure 8 from \cite{matthews1969sensitivity}).  The amplitude of the square pulse was matched to the amplitude of sinusoidal stimulation (following the usual definition of sinusoidal amplitude).  The square pulse response is useful in illustrating how the mechanism of adaptation unfolds theoretically.  From (\ref{gut1}) and (\ref{gut2}) we observe that the firing rate is essentially a monotonic function of intensity divided by sample size.  Intensity can change abruptly but sample size must remain continuous (due to the continuity of $m$).  When the stimulus is turned on, the ratio of $I$ to $m$ becomes large but falls as sample size grows to match the input when $m$ approaches $m_{eq}$.  At the termination of the input, the ratio falls abruptly to a small value before returning to steady-state as sample size decreases to match the input.  Such behaviour is typical of adaptation/de-adaptation responses and we can now observe this mathematically.

\begin{figure}
\centering
\includegraphics[width=0.45\textwidth]{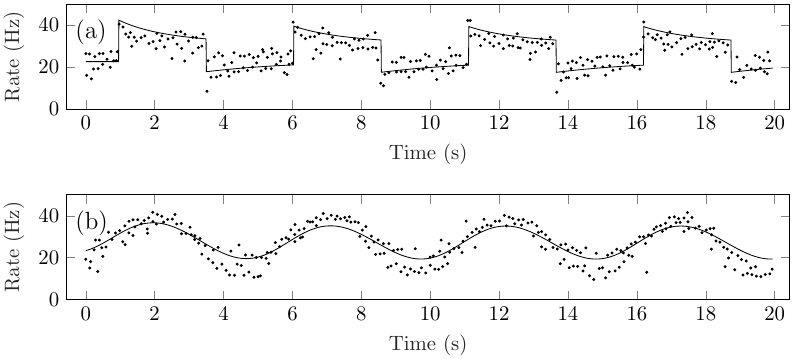}
\caption{The response of a cat muscle spindle to repeated stimulation \cite{matthews1969sensitivity}.  (a) Response to square pulse stimulation compared to (b) sinusoidal stimulation.  The amplitude for both inputs was identical (50 $\mu$m stretch).  Four parameters were used for both experiments from a numerical solution of (\ref{gut1}-\ref{gut4}).}
\label{fig:fig8}
\end{figure}

The challenge in using (\ref{gut1}-\ref{gut4}) lies in estimating the five unknown parameters.  The equations themselves are robust in that a wide range of parameter values will give the correct qualitative shape of response.  However determining values which yield the global optimal solution (or even the initial values for a non-linear curve-fit) is difficult when a numerical solution to the differential equation is required.  Next, we illustrate several techniques that can be used to provide an estimate when an adaptation response is provided.  In the small intensity limit, a Taylor series expansion of (\ref{adapt}) allows for $k$ and $\beta$ to be combined into a new single parameter.  Solving the analytical response to a step input, we use the mathematical form of the solution to solve for the steady-state response prior to stimulus onset SR and the subsequent new steady-state SS.  The inverse time-constant $a$ can be calculated from $\text{SR}=22$ Hz, $\text{SS}=30$ Hz and one additional point from the data (firing rate response of 33 Hz at $t=13$ s from Figure \ref{fig:fig8}a).  From here, the choice of SS and SR constrains the equations such that only two adjustable parameters remain.  Choosing $p=1.0$ yields $k\beta=6.0$ and $\delta I=57$.  Together with $a=0.51$ Hz, the same parameters were used to obtain the sinusoidal response shown in Figure \ref{fig:fig8}b without the need to curve-fit.  

\subsubsection{Response amplitude as a function of modulation frequency in retinal ganglion cells}
When light intensity is varied sinusoidally, the resulting neural response will be periodic.  Figure \ref{fig:fig1}d shows an example of a sinusoidally modulated input.  The modulation depth is defined as $\left(I_{\text{max}}-I_{\text{min}}\right)/\left(I_{\text{max}}+I_{\text{min}}\right)$.  When this index is small, the equations can be solved analytically for an input of the form $I+\Delta I \sin\left(\omega t\right)$.  The solution will have both a transient and a steady-state component.  We are interested in the steady-state solution.

We begin by defining $Y=\Delta I/(I+\delta I)$.  $Y$ is equal to the modulation depth when $\delta I=0$.  In the limit of small $Y$, a linear expansion gives
\begin{multline}
F_{SS}=\frac{1}{2} k\log\left(1+\beta \left(I+\delta I\right)^{p/2}\right) \\ 
+ kY C_1(I) C_2(\omega) \sin(\omega t+\phi) \label{sinusoidal}
\end{multline}
where 
\begin{align}
&C_1(I)=\frac{1}{2}\frac{\beta\left(I+\delta I\right)^{p/2} }{1+\beta\left(I+\delta I\right)^{p/2}} \\
&C_2(\omega)=\frac{p}{2}\sqrt{\frac{1+4\omega^2/a^2}{1+\omega^2/a^2}} \\
&\phi=\arctan\left(\frac{\omega/a}{1+2\omega^2/a^2}\right)
\end{align}
That is, the steady-state response is itself sinusoidal.  

\begin{figure}
\centering
\includegraphics[width=0.25\textwidth]{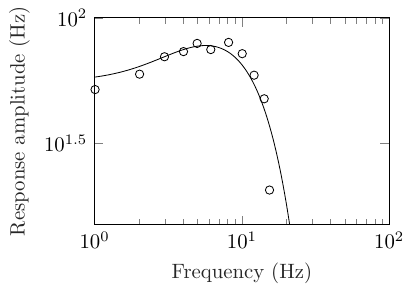}
\caption{Response amplitude of a cat retinal ganglion cell as a function of modulation frequency for a sinusoidally modulated input (open circles) \cite{cleland1966cat}.  The smooth curve generated from the theory was fitted to the data using only a horizontal and a vertical shift in the log-log domain.  Please see text for more details.}
\label{fig:fig9}
\end{figure}

Figure \ref{fig:fig9} shows the response of a cat retinal ganglion cell to sinusoidally modulated light (from Figure 9a of \cite{cleland1966cat}).  Response amplitude is defined as the difference between the highest and lowest firing rates, and was measured as a function of modulation frequency.  Before comparing (\ref{sinusoidal}) to data, it is important to remember that the response of a single ganglion cell is determined from the input of many photoreceptor cells.  Following \cite{enroth1966contrast}, it has been shown that these individual photoreceptor inputs sum linearly.  However, due to differences in time of arrival, a jitter is introduced.  By the central limit theorem, the jitter can be considered normally distributed.  The amplitude of the average response from the ganglion cell is therefore convolved with a Gaussian kernel with zero mean and variance $\sigma_\text{jitter}^2$.  From here the average response amplitude can be calculated to be
\begin{equation}
2 kY C_1(\omega) C_2(I) \exp\left(-\omega^2\sigma_\text{jitter}^2/2\right)
\end{equation}
where there is now a frequency-dependent drop-off in the response amplitude.  Other studies have speculated on the origins of the low-pass characteristics of the neural response in terms of membrane time constants and average firing rates, e.g. \cite{knight1972dynamics,benda2008spike}.

Finally, using the time-frequency bandwidth tradeoff, the width of the time jitter can be related to the time constant of adaptation so that $\sigma_\text{jitter}=1/2a$.  Thus, response amplitude becomes
\begin{equation}
2kY C_1(\omega) C_2(I) \exp\left(-\omega^2/8a^2\right) \label{flicker}
\end{equation}
In the large intensity limit, this is an equation of four parameters which can be fitted to experimental data.  In Figure \ref{fig:fig9} the response amplitude was measured to a signal with modulation depth of one-half \cite{cleland1966cat}.  This likely violates the condition under which (\ref{sinusoidal}) was derived.  Nevertheless, a comparison of theory to data was attempted.  Since intensity is fixed, there are only two fitting parameters: one is the scaling factor for response amplitude, the other is $a$ which scales the frequency axis as the equation is solely a function of $\omega/a$.  On a log-log plot, this amounts to a vertical or horizontal shift.  Everything else is `locked in' by the theory.  Equation (\ref{flicker}) is plotted along side the data in Figure \ref{fig:fig9}.  The entire characteristic shape of the response curve is reproduced, including the inflection observed at low frequencies.

\section{Prediction of a new law and universality} 
\label{section:new}
The value of a theory is not only to provide an account of existing data, but also to predict hitherto unobserved phenomena, patterns and behaviour in nature.  As such, this final section is devoted to the derivation of a yet undiscovered sensory law that is obeyed as far back as the original measurements of Edgar Adrian in his discovery of the all-or-nothing principle of action potentials.

This law can be derived through the small intensity approximation of (\ref{gut2}) together with (\ref{gut1}).  In the case where $\beta \left(I+\delta I\right)^p/m \ll 1$, we take a first-order approximation to obtain $F = \left(I+\delta I\right)^p/m$.  Without loss of generality, we have assumed $k\beta/2=1$.  The value of these constants are not germane to the discussion.  Evaluating the expression for the spontaneous SR, peak PR and steady-state SS values for adaptation with (\ref{begin}-\ref{end}) we have
\begin{align}
&\text{SR}= \delta I^{p/2} \\
&\text{PR}= \dfrac{\left(I+\delta I\right)^p}{\delta I^{p/2}} \\
&\text{SS}= \left(I+\delta I\right)^{p/2}
\end{align}
Finally, these equations can be combined into a single succinct expression: 
\begin{equation}
\label{golden}
\text{SS}=\sqrt{\text{PR}\times \text{SR}}
\end{equation}

Equation (\ref{golden}) is free of any elements of the theory and requires no parameters to be estimated.  As an equation of biology, it is an expression of remarkable simplicity and elegance --  a new and undiscovered equation of neurophysiology!  In plain terms, (\ref{golden}) states that, under perturbation, the final steady-state response is equal to the geometric mean of the initial and peak responses.  As is documented in \cite{Wong2020.02.17.953448}, (\ref{golden}) is compatible with measurements of adaptation as far back as the original recordings in the frog muscle by Adrian and Zotterman \cite{adrian1926impulsesb}.  It is also compatible with data from a wide range of modalities, over a wide variety of organisms.  This has several important implications.  First, since it was derived from the theory, any set of adaptation measurements obeying (\ref{golden}) will likely be well-described by (\ref{gut1}-\ref{gut4}) \textit{over the entire time course of adaptation without even the need to attempt a fit.}  Second, the ubiquity of (\ref{golden}) would suggest that the approach detailed here is universal in its description of sensory function, comprising the principles by which all living things are governed.

The approach detailed in this paper is free of consideration of the mechanism of transduction, focusing instead on the principles that underlie the processing of information in intensity coding.  Equation (\ref{golden}) exemplifies the merits of this approach as it is obeyed across a wide range of modalities and organism species including at least one organism from each major phylum in \textit{Animalia} \cite{Wong2020.02.17.953448}.  Borrowing from the concept of consilience, when diverse, unrelated studies across different animal species converge on the same result, this is powerful evidence of a common theory of sensory processing.  However, the absence of discussion on mechanism does not mean that consideration of the underlying biophysics is unimportant.  In fact, the two approaches go hand-in-hand in furthering our understanding of the senses.  For example, spontaneous activity is often thought of as noise in the nervous system \cite{imaizumi2018spontaneous}.  However, it is clear from (\ref{golden}) that spontaneous activity plays an integral part in normal sensory function irrespective of the mechanism of generation.  Theory can help guide the investigation of sensory mechanisms.

Despite its success, however, the theory does not work well with all modalities.  Why is this the case and what are some of the issues?  First, discrepancies have been found in modalities belonging to the category of ``active sensory systems''.  These are systems which sense the environment through the use of self-generated signals and requires the receptors to distinguish between internal and external inputs (i.e. the concept of \textit{reafference} \cite{von1950reafferenzprinzip}).  Such modalities would include thermoreception (where transduction depends on whether the stimulus is hotter or colder than the body temperature), electroreception (using self-generated electric fields to sense distortions to this field) and the vestibular system.  Active sensory systems may require considerations which differ from purely passive senses like vision.  For the theory, this would involve rethinking the internal signal component $\delta I$.  Other issues include the amount of adaptation exhibited by the unit.  Tonic units, for example, adapt more slowly and show a sustained response in contrast to phasic units which adapt quickly and stop firing.  But the reasons which lead to a difference in response can be more subtle.  For example, it has been proposed that the phasic response in touch may in fact depend on the time derivative of the input rather than the input itself \cite{kim2010predicting}.  As such, the difficulty of replicating such responses theoretically may come down to a better understanding of the nature of the input into the system.

\section{Final remarks}
The theory developed in this paper has a particular mathematical simplicity because we have restricted the analysis to the asymptotic or large intensity, near-equilibrium or near steady-state limit.  The situation is more difficult if we considered the non-equilibrium case (small $m$, far from $m_{eq}$).  In such cases, the response may depend strongly on the prior distribution $\pi_0(\theta)$ from (\ref{bayes}) or on the precise mathematical form of the sampling rate function $dm/dt$ in (\ref{sample}).  Despite this, we have shown that the equations hold predictive power across different time scales, for almost all sensory modalities and different animal species.

There is however one fundamental result governing the sensory response that can be derived even when far from equilibrium.  Using a basic theorem of information theory \cite{cover2012elements}, we can write for the entropy of the posterior distribution
\begin{equation}
H(\theta|X_1,...,X_m) \le H(\theta)
\end{equation}
where $H(\theta)$ is the entropy of the prior distribution and $H(\theta|X_1,...,X_m)$ the entropy over the posterior distribution.  That is, entropy decreases or remains constant with additional samples or measurements.  Since $F=kH$ and $dm/dt \ge 0$, we have
\begin{equation}
dF/dt \le 0
\end{equation}
 \textit{Thus, with minimal assumptions, we have proved that the sensory response to a constant stimulus must, on average, decrease or remain constant.}  This inequality, together with the use of Boltzmann-Shannon entropy and $F=kH$ suggests a deeper connection between sensory processing and statistical physics.

\section{Acknowledgements}
This work was supported by a Discovery Grant from the Natural Sciences and Engineering Research Council of Canada (NSERC).  The author is grateful for the many helpful discussions with Professor Kenneth Norwich, continued encouragement from Professor Manfredi Maggiore, and those who have contributed to the entropy theory in the past: Sheldon Opps, Suraya Figueiredo, Gerry Fung, Zoe Zhao, Bruno de Oliveira Floriano, Prathima Sundaram and Sai Vemula. 

This is a preprint of an article published in Biological Cybernetics. The final authenticated version is available online at: https://doi.org/10.1007/s00422-020-00848-4

\appendix
\section{The optimal sample size}
\label{section:optimal}
The optimal sample size $m_{eq}$ is the number of samples after which $m$ no longer changes. Since the fluctuation scaling law posits that the variance of a signal increases with the magnitude of the signal, a choice of constant $m_{eq}$ implies that the estimation error in the mean will increase when intensity is increased.  On the other hand, if the standard error $\sqrt{\sigma^2/m_{eq}}$ is held constant, $m_{eq}$ must then take the form
\begin{equation}
m_{eq} \propto \left(I+\delta I\right)^p \label{meq^2}
\end{equation}
requiring sample size to change significantly with intensity.

Between these two extremes lies a third possibility.  Consider the situation where the sensory system is presented with an input of intensity $I_1$ which is later changed to $I_2$.  Without loss of generality, assume that $I_2>I_1$.  At steady-state, the standard error of $I_1$ is $\text{SE}_1$.  Let the initial uncertainty in $I_2$ be $\text{SE}_\text{2,initial}$.  Increasing the number of samples will cause this error to fall.  How is the steady-state error in $I_2$ determined?  Taking $\text{SE}_\text{2}$ to be the geometric average of the standard errors, we obtain
\begin{equation}
\text{SE}_2=\left(\text{SE}_1 \times \text{SE}_{\text{2,initial}} \right)^{1/2} \label{fisherinfo1}
\end{equation}
That is, the error in estimating $I_2$ is equal to the average of the steady-state error in $I_1$ and the initial uncertainty in $I_2$.

To see what effect (\ref{fisherinfo1}) has on the optimal sample size, let $m_2(0)$ be the initial sample size just after the change in intensity.  Since $m$ must remain continuous across the boundary we have $m_2(0)=m_{eq,1}$, where $m_{eq,1}$ is the optimal sample size for $I_1$.  This calculation assumes that steady-state is achieved prior to the change in intensity.  From this, we conclude that the expression $\sigma^2/m_{eq}^2$ is invariant to changes in intensity.  Thus the general relationship between optimal sample size and intensity is 
\begin{equation}
m_{eq} =c (I+\delta I)^{p/2} \label{meq}
\end{equation}
where $c$ is a constant.  For simplicity, $c$ is set to unity as it can be incorporated into $\beta$ in (\ref{gut2}).  Equation (\ref{meq}) is the expression for optimal sample size used in the theory and can be tested both directly and indirectly through comparison with experimental data.

\section{Key assumptions of the theory}
\label{section:assumptions}
\begin{enumerate}
\item The sensory receptor draws repeated, independent samples of the stimulus magnitude to estimate the mean.
\item Samples are processed with limited resolution resulting in normally distributed error with zero mean and variance $R$, which is constant relative to the input.
\item Firing rate is proportional to the Boltzmann-Shannon measure of uncertainty in the mean.
\item The statistics of the sensory signal are governed by a Tweedie distribution.
\item The signal mean $\mu$ is a sum of the experimenter controlled intensity $I$ plus constant additive background noise $\delta I$.
\item Sampling rate is a function of the difference between the current and optimal sample sizes.
\item The optimal sample size is determined from an average of standard errors in the mean.
\end{enumerate}

\bibliography{asymptotic}

\begin{thebibliography}{41}%
\makeatletter
\providecommand \@ifxundefined [1]{%
 \@ifx{#1\undefined}
}%
\providecommand \@ifnum [1]{%
 \ifnum #1\expandafter \@firstoftwo
 \else \expandafter \@secondoftwo
 \fi
}%
\providecommand \@ifx [1]{%
 \ifx #1\expandafter \@firstoftwo
 \else \expandafter \@secondoftwo
 \fi
}%
\providecommand \natexlab [1]{#1}%
\providecommand \enquote  [1]{``#1''}%
\providecommand \bibnamefont  [1]{#1}%
\providecommand \bibfnamefont [1]{#1}%
\providecommand \citenamefont [1]{#1}%
\providecommand \href@noop [0]{\@secondoftwo}%
\providecommand \href [0]{\begingroup \@sanitize@url \@href}%
\providecommand \@href[1]{\@@startlink{#1}\@@href}%
\providecommand \@@href[1]{\endgroup#1\@@endlink}%
\providecommand \@sanitize@url [0]{\catcode `\\12\catcode `\$12\catcode
  `\&12\catcode `\#12\catcode `\^12\catcode `\_12\catcode `\%12\relax}%
\providecommand \@@startlink[1]{}%
\providecommand \@@endlink[0]{}%
\providecommand \url  [0]{\begingroup\@sanitize@url \@url }%
\providecommand \@url [1]{\endgroup\@href {#1}{\urlprefix }}%
\providecommand \urlprefix  [0]{URL }%
\providecommand \Eprint [0]{\href }%
\providecommand \doibase [0]{http://dx.doi.org/}%
\providecommand \selectlanguage [0]{\@gobble}%
\providecommand \bibinfo  [0]{\@secondoftwo}%
\providecommand \bibfield  [0]{\@secondoftwo}%
\providecommand \translation [1]{[#1]}%
\providecommand \BibitemOpen [0]{}%
\providecommand \bibitemStop [0]{}%
\providecommand \bibitemNoStop [0]{.\EOS\space}%
\providecommand \EOS [0]{\spacefactor3000\relax}%
\providecommand \BibitemShut  [1]{\csname bibitem#1\endcsname}%
\let\auto@bib@innerbib\@empty
\bibitem [{\citenamefont {Norwich}(1977)}]{norwich1977information}%
  \BibitemOpen
  \bibfield  {author} {\bibinfo {author} {\bibfnamefont {K.~H.}\ \bibnamefont
  {Norwich}},\ }\href@noop {} {\bibfield  {journal} {\bibinfo  {journal}
  {Bulletin of Mathematical Biology}\ }\textbf {\bibinfo {volume} {39}},\
  \bibinfo {pages} {453} (\bibinfo {year} {1977})}\BibitemShut {NoStop}%
\bibitem [{\citenamefont {Norwich}(1993)}]{norwich1993information}%
  \BibitemOpen
  \bibfield  {author} {\bibinfo {author} {\bibfnamefont {K.~H.}\ \bibnamefont
  {Norwich}},\ }\href@noop {} {\emph {\bibinfo {title} {Information, Sensation,
  and Perception}}}\ (\bibinfo  {publisher} {Academic Press San Diego},\
  \bibinfo {year} {1993})\BibitemShut {NoStop}%
\bibitem [{\citenamefont {Norwich}\ and\ \citenamefont
  {Wong}(1995)}]{norwich1995universal}%
  \BibitemOpen
  \bibfield  {author} {\bibinfo {author} {\bibfnamefont {K.~H.}\ \bibnamefont
  {Norwich}}\ and\ \bibinfo {author} {\bibfnamefont {W.}~\bibnamefont {Wong}},\
  }\href@noop {} {\bibfield  {journal} {\bibinfo  {journal} {Mathematical
  Biosciences}\ }\textbf {\bibinfo {volume} {125}},\ \bibinfo {pages} {83}
  (\bibinfo {year} {1995})}\BibitemShut {NoStop}%
\bibitem [{\citenamefont {Wong}(1997)}]{wong1997physics}%
  \BibitemOpen
  \bibfield  {author} {\bibinfo {author} {\bibfnamefont {W.}~\bibnamefont
  {Wong}},\ }\emph {\bibinfo {title} {On the Physics of Perception}},\
  \href@noop {} {Ph.D. thesis},\ \bibinfo  {school} {University of Toronto}
  (\bibinfo {year} {1997})\BibitemShut {NoStop}%
\bibitem [{\citenamefont {Benda}\ and\ \citenamefont
  {Herz}(2003)}]{benda2003universal}%
  \BibitemOpen
  \bibfield  {author} {\bibinfo {author} {\bibfnamefont {J.}~\bibnamefont
  {Benda}}\ and\ \bibinfo {author} {\bibfnamefont {A.~V.}\ \bibnamefont
  {Herz}},\ }\href@noop {} {\bibfield  {journal} {\bibinfo  {journal} {Neural
  Computation}\ }\textbf {\bibinfo {volume} {15}},\ \bibinfo {pages} {2523}
  (\bibinfo {year} {2003})}\BibitemShut {NoStop}%
\bibitem [{\citenamefont {Drew}\ and\ \citenamefont
  {Abbott}(2006)}]{drew2006models}%
  \BibitemOpen
  \bibfield  {author} {\bibinfo {author} {\bibfnamefont {P.~J.}\ \bibnamefont
  {Drew}}\ and\ \bibinfo {author} {\bibfnamefont {L.~F.}\ \bibnamefont
  {Abbott}},\ }\href@noop {} {\bibfield  {journal} {\bibinfo  {journal}
  {Journal of Neurophysiology}\ }\textbf {\bibinfo {volume} {96}},\ \bibinfo
  {pages} {826} (\bibinfo {year} {2006})}\BibitemShut {NoStop}%
\bibitem [{\citenamefont {Aviel}\ and\ \citenamefont
  {Gerstner}(2006)}]{aviel2006spiking}%
  \BibitemOpen
  \bibfield  {author} {\bibinfo {author} {\bibfnamefont {Y.}~\bibnamefont
  {Aviel}}\ and\ \bibinfo {author} {\bibfnamefont {W.}~\bibnamefont
  {Gerstner}},\ }\href@noop {} {\bibfield  {journal} {\bibinfo  {journal}
  {Physical Review E}\ }\textbf {\bibinfo {volume} {73}},\ \bibinfo {pages}
  {051908} (\bibinfo {year} {2006})}\BibitemShut {NoStop}%
\bibitem [{\citenamefont {Famulare}\ and\ \citenamefont
  {Fairhall}(2010)}]{famulare2010feature}%
  \BibitemOpen
  \bibfield  {author} {\bibinfo {author} {\bibfnamefont {M.}~\bibnamefont
  {Famulare}}\ and\ \bibinfo {author} {\bibfnamefont {A.}~\bibnamefont
  {Fairhall}},\ }\href@noop {} {\bibfield  {journal} {\bibinfo  {journal}
  {Neural Computation}\ }\textbf {\bibinfo {volume} {22}},\ \bibinfo {pages}
  {581} (\bibinfo {year} {2010})}\BibitemShut {NoStop}%
\bibitem [{\citenamefont {Van~der Vaart}(2000)}]{van2000asymptotic}%
  \BibitemOpen
  \bibfield  {author} {\bibinfo {author} {\bibfnamefont {A.~W.}\ \bibnamefont
  {Van~der Vaart}},\ }\href@noop {} {\emph {\bibinfo {title} {Asymptotic
  Statistics}}}\ (\bibinfo  {publisher} {Cambridge university press},\ \bibinfo
  {year} {2000})\BibitemShut {NoStop}%
\bibitem [{\citenamefont {Cover}\ and\ \citenamefont
  {Thomas}(2012)}]{cover2012elements}%
  \BibitemOpen
  \bibfield  {author} {\bibinfo {author} {\bibfnamefont {T.~M.}\ \bibnamefont
  {Cover}}\ and\ \bibinfo {author} {\bibfnamefont {J.~A.}\ \bibnamefont
  {Thomas}},\ }\href@noop {} {\emph {\bibinfo {title} {Elements of Information
  Theory}}}\ (\bibinfo  {publisher} {John Wiley \& Sons, New York},\ \bibinfo
  {year} {2012})\BibitemShut {NoStop}%
\bibitem [{\citenamefont {Doksum}\ and\ \citenamefont
  {Bickel}(2007)}]{doksum2007mathematical}%
  \BibitemOpen
  \bibfield  {author} {\bibinfo {author} {\bibfnamefont {K.~A.}\ \bibnamefont
  {Doksum}}\ and\ \bibinfo {author} {\bibfnamefont {P.}~\bibnamefont
  {Bickel}},\ }\href@noop {} {\emph {\bibinfo {title} {Mathematical Statistics:
  Basic Ideas and Selected Topics}}}\ (\bibinfo  {publisher} {Prentice Hall},\
  \bibinfo {year} {2007})\BibitemShut {NoStop}%
\bibitem [{\citenamefont {Norwich}(1983)}]{norwich1983perceive}%
  \BibitemOpen
  \bibfield  {author} {\bibinfo {author} {\bibfnamefont {K.~H.}\ \bibnamefont
  {Norwich}},\ }\href@noop {} {\bibfield  {journal} {\bibinfo  {journal}
  {Journal of Theoretical Biology}\ }\textbf {\bibinfo {volume} {102}},\
  \bibinfo {pages} {175} (\bibinfo {year} {1983})}\BibitemShut {NoStop}%
\bibitem [{\citenamefont {Kostal}\ and\ \citenamefont
  {D'Onofrio}(2018)}]{kostal2018coordinate}%
  \BibitemOpen
  \bibfield  {author} {\bibinfo {author} {\bibfnamefont {L.}~\bibnamefont
  {Kostal}}\ and\ \bibinfo {author} {\bibfnamefont {G.}~\bibnamefont
  {D'Onofrio}},\ }\href@noop {} {\bibfield  {journal} {\bibinfo  {journal}
  {Biological Cybernetics}\ }\textbf {\bibinfo {volume} {112}},\ \bibinfo
  {pages} {13} (\bibinfo {year} {2018})}\BibitemShut {NoStop}%
\bibitem [{\citenamefont {Marks}(1974)}]{marks1974sensory}%
  \BibitemOpen
  \bibfield  {author} {\bibinfo {author} {\bibfnamefont {L.}~\bibnamefont
  {Marks}},\ }\href@noop {} {\emph {\bibinfo {title} {Sensory Processes: The
  New Psychophysics}}}\ (\bibinfo  {publisher} {Elsevier},\ \bibinfo {year}
  {1974})\BibitemShut {NoStop}%
\bibitem [{\citenamefont {Ratliff}\ and\ \citenamefont
  {Riggs}(1950)}]{ratliff1950involuntary}%
  \BibitemOpen
  \bibfield  {author} {\bibinfo {author} {\bibfnamefont {F.}~\bibnamefont
  {Ratliff}}\ and\ \bibinfo {author} {\bibfnamefont {L.~A.}\ \bibnamefont
  {Riggs}},\ }\href@noop {} {\bibfield  {journal} {\bibinfo  {journal} {Journal
  of Experimental Psychology}\ }\textbf {\bibinfo {volume} {40}},\ \bibinfo
  {pages} {687} (\bibinfo {year} {1950})}\BibitemShut {NoStop}%
\bibitem [{\citenamefont {Laming}(1986)}]{laming1986sensory}%
  \BibitemOpen
  \bibfield  {author} {\bibinfo {author} {\bibfnamefont {D.}~\bibnamefont
  {Laming}},\ }\href@noop {} {\emph {\bibinfo {title} {Sensory Analysis}}}\
  (\bibinfo  {publisher} {Cambridge University Press, Cambridge},\ \bibinfo
  {year} {1986})\BibitemShut {NoStop}%
\bibitem [{\citenamefont {Itti}\ and\ \citenamefont
  {Baldi}(2009)}]{itti2009bayesian}%
  \BibitemOpen
  \bibfield  {author} {\bibinfo {author} {\bibfnamefont {L.}~\bibnamefont
  {Itti}}\ and\ \bibinfo {author} {\bibfnamefont {P.}~\bibnamefont {Baldi}},\
  }\href@noop {} {\bibfield  {journal} {\bibinfo  {journal} {Vision Research}\
  }\textbf {\bibinfo {volume} {49}},\ \bibinfo {pages} {1295} (\bibinfo {year}
  {2009})}\BibitemShut {NoStop}%
\bibitem [{\citenamefont {Paul}(1982)}]{paul1982photon}%
  \BibitemOpen
  \bibfield  {author} {\bibinfo {author} {\bibfnamefont {H.}~\bibnamefont
  {Paul}},\ }\href@noop {} {\bibfield  {journal} {\bibinfo  {journal} {Reviews
  of Modern Physics}\ }\textbf {\bibinfo {volume} {54}},\ \bibinfo {pages}
  {1061} (\bibinfo {year} {1982})}\BibitemShut {NoStop}%
\bibitem [{\citenamefont {Adrian}\ and\ \citenamefont
  {Zotterman}(1926)}]{adrian1926impulsesb}%
  \BibitemOpen
  \bibfield  {author} {\bibinfo {author} {\bibfnamefont {E.~D.}\ \bibnamefont
  {Adrian}}\ and\ \bibinfo {author} {\bibfnamefont {Y.}~\bibnamefont
  {Zotterman}},\ }\href@noop {} {\bibfield  {journal} {\bibinfo  {journal}
  {Journal of Physiology}\ }\textbf {\bibinfo {volume} {61}},\ \bibinfo {pages}
  {151} (\bibinfo {year} {1926})}\BibitemShut {NoStop}%
\bibitem [{\citenamefont {Taylor}(1961)}]{taylor1961aggregation}%
  \BibitemOpen
  \bibfield  {author} {\bibinfo {author} {\bibfnamefont {L.}~\bibnamefont
  {Taylor}},\ }\href@noop {} {\bibfield  {journal} {\bibinfo  {journal}
  {Nature}\ }\textbf {\bibinfo {volume} {189}},\ \bibinfo {pages} {732}
  (\bibinfo {year} {1961})}\BibitemShut {NoStop}%
\bibitem [{\citenamefont {Kendal}\ and\ \citenamefont
  {J{\o}rgensen}(2011)}]{kendal2011taylor}%
  \BibitemOpen
  \bibfield  {author} {\bibinfo {author} {\bibfnamefont {W.~S.}\ \bibnamefont
  {Kendal}}\ and\ \bibinfo {author} {\bibfnamefont {B.}~\bibnamefont
  {J{\o}rgensen}},\ }\href@noop {} {\bibfield  {journal} {\bibinfo  {journal}
  {Physical Review E}\ }\textbf {\bibinfo {volume} {83}},\ \bibinfo {pages}
  {066115} (\bibinfo {year} {2011})}\BibitemShut {NoStop}%
\bibitem [{\citenamefont {J{\o}rgensen}(1997)}]{jorgensen1997theory}%
  \BibitemOpen
  \bibfield  {author} {\bibinfo {author} {\bibfnamefont {B.}~\bibnamefont
  {J{\o}rgensen}},\ }\href@noop {} {\emph {\bibinfo {title} {The Theory of
  Dispersion Models}}}\ (\bibinfo  {publisher} {Chapman \& Hall, London},\
  \bibinfo {year} {1997})\BibitemShut {NoStop}%
\bibitem [{\citenamefont {Kuhn}\ \emph {et~al.}(2004)\citenamefont {Kuhn},
  \citenamefont {Aertsen},\ and\ \citenamefont {Rotter}}]{kuhn2004neuronal}%
  \BibitemOpen
  \bibfield  {author} {\bibinfo {author} {\bibfnamefont {A.}~\bibnamefont
  {Kuhn}}, \bibinfo {author} {\bibfnamefont {A.}~\bibnamefont {Aertsen}}, \
  and\ \bibinfo {author} {\bibfnamefont {S.}~\bibnamefont {Rotter}},\
  }\href@noop {} {\bibfield  {journal} {\bibinfo  {journal} {Journal of
  Neuroscience}\ }\textbf {\bibinfo {volume} {24}},\ \bibinfo {pages} {2345}
  (\bibinfo {year} {2004})}\BibitemShut {NoStop}%
\bibitem [{\citenamefont {Schwalger}\ \emph {et~al.}(2010)\citenamefont
  {Schwalger}, \citenamefont {Fisch}, \citenamefont {Benda},\ and\
  \citenamefont {Lindner}}]{schwalger2010noisy}%
  \BibitemOpen
  \bibfield  {author} {\bibinfo {author} {\bibfnamefont {T.}~\bibnamefont
  {Schwalger}}, \bibinfo {author} {\bibfnamefont {K.}~\bibnamefont {Fisch}},
  \bibinfo {author} {\bibfnamefont {J.}~\bibnamefont {Benda}}, \ and\ \bibinfo
  {author} {\bibfnamefont {B.}~\bibnamefont {Lindner}},\ }\href@noop {}
  {\bibfield  {journal} {\bibinfo  {journal} {PLoS Computational Biology}\
  }\textbf {\bibinfo {volume} {6}},\ \bibinfo {pages} {e1001026} (\bibinfo
  {year} {2010})}\BibitemShut {NoStop}%
\bibitem [{\citenamefont {Pifferi}\ \emph {et~al.}(2009)\citenamefont
  {Pifferi}, \citenamefont {Menini},\ and\ \citenamefont
  {Kurahashi}}]{pifferi2009signal}%
  \BibitemOpen
  \bibfield  {author} {\bibinfo {author} {\bibfnamefont {S.}~\bibnamefont
  {Pifferi}}, \bibinfo {author} {\bibfnamefont {A.}~\bibnamefont {Menini}}, \
  and\ \bibinfo {author} {\bibfnamefont {T.}~\bibnamefont {Kurahashi}},\ }in\
  \href@noop {} {\emph {\bibinfo {booktitle} {The Neurobiology of
  Olfaction}}},\ \bibinfo {editor} {edited by\ \bibinfo {editor} {\bibfnamefont
  {A.}~\bibnamefont {Menini}}}\ (\bibinfo  {publisher} {CRC Press, Boca Raton
  FL, USA},\ \bibinfo {year} {2009})\ pp.\ \bibinfo {pages}
  {203--224}\BibitemShut {NoStop}%
\bibitem [{\citenamefont {Smith}(1988)}]{smith1988encoding}%
  \BibitemOpen
  \bibfield  {author} {\bibinfo {author} {\bibfnamefont {R.~L.}\ \bibnamefont
  {Smith}},\ }in\ \href@noop {} {\emph {\bibinfo {booktitle} {Auditory
  Function: Neurobiological Bases of Hearing}}},\ \bibinfo {editor} {edited by\
  \bibinfo {editor} {\bibfnamefont {G.~M.}\ \bibnamefont {Edelman}}, \bibinfo
  {editor} {\bibfnamefont {W.~E.}\ \bibnamefont {Gall}}, \ and\ \bibinfo
  {editor} {\bibfnamefont {W.~M.}\ \bibnamefont {Cowan}}}\ (\bibinfo
  {publisher} {John Wiley \& Sons, Toronto, Canada},\ \bibinfo {year} {1988})\
  pp.\ \bibinfo {pages} {243--274}\BibitemShut {NoStop}%
\bibitem [{\citenamefont {Smith}\ and\ \citenamefont
  {Zwislocki}(1975)}]{smith1975short}%
  \BibitemOpen
  \bibfield  {author} {\bibinfo {author} {\bibfnamefont {R.~L.}\ \bibnamefont
  {Smith}}\ and\ \bibinfo {author} {\bibfnamefont {J.}~\bibnamefont
  {Zwislocki}},\ }\href@noop {} {\bibfield  {journal} {\bibinfo  {journal}
  {Biological Cybernetics}\ }\textbf {\bibinfo {volume} {17}},\ \bibinfo
  {pages} {169} (\bibinfo {year} {1975})}\BibitemShut {NoStop}%
\bibitem [{\citenamefont {Benda}\ \emph {et~al.}(2005)\citenamefont {Benda},
  \citenamefont {Longtin},\ and\ \citenamefont {Maler}}]{benda2005spike}%
  \BibitemOpen
  \bibfield  {author} {\bibinfo {author} {\bibfnamefont {J.}~\bibnamefont
  {Benda}}, \bibinfo {author} {\bibfnamefont {A.}~\bibnamefont {Longtin}}, \
  and\ \bibinfo {author} {\bibfnamefont {L.}~\bibnamefont {Maler}},\
  }\href@noop {} {\bibfield  {journal} {\bibinfo  {journal} {Journal of
  Neuroscience}\ }\textbf {\bibinfo {volume} {25}},\ \bibinfo {pages} {2312}
  (\bibinfo {year} {2005})}\BibitemShut {NoStop}%
\bibitem [{\citenamefont {Dethier}\ and\ \citenamefont
  {Bowdan}(1984)}]{dethier1984relations}%
  \BibitemOpen
  \bibfield  {author} {\bibinfo {author} {\bibfnamefont {V.}~\bibnamefont
  {Dethier}}\ and\ \bibinfo {author} {\bibfnamefont {E.}~\bibnamefont
  {Bowdan}},\ }\href@noop {} {\bibfield  {journal} {\bibinfo  {journal}
  {Behavioral Neuroscience}\ }\textbf {\bibinfo {volume} {98}},\ \bibinfo
  {pages} {791} (\bibinfo {year} {1984})}\BibitemShut {NoStop}%
\bibitem [{\citenamefont {Norwich}\ and\ \citenamefont
  {McConville}(1991)}]{norwich1991informational}%
  \BibitemOpen
  \bibfield  {author} {\bibinfo {author} {\bibfnamefont {K.~H.}\ \bibnamefont
  {Norwich}}\ and\ \bibinfo {author} {\bibfnamefont {K.~M.~V.}\ \bibnamefont
  {McConville}},\ }\href@noop {} {\bibfield  {journal} {\bibinfo  {journal}
  {Journal of Comparative Physiology A}\ }\textbf {\bibinfo {volume} {168}},\
  \bibinfo {pages} {151} (\bibinfo {year} {1991})}\BibitemShut {NoStop}%
\bibitem [{\citenamefont {Sch{\"a}fer}(1994)}]{schafer1994regularity}%
  \BibitemOpen
  \bibfield  {author} {\bibinfo {author} {\bibfnamefont {S.}~\bibnamefont
  {Sch{\"a}fer}},\ }\href@noop {} {\bibfield  {journal} {\bibinfo  {journal}
  {Experimental Brain Research}\ }\textbf {\bibinfo {volume} {102}},\ \bibinfo
  {pages} {198} (\bibinfo {year} {1994})}\BibitemShut {NoStop}%
\bibitem [{\citenamefont {Bohnenberger}(1981)}]{bohnenberger1981matched}%
  \BibitemOpen
  \bibfield  {author} {\bibinfo {author} {\bibfnamefont {J.}~\bibnamefont
  {Bohnenberger}},\ }\href@noop {} {\bibfield  {journal} {\bibinfo  {journal}
  {Journal of Comparative Physiology}\ }\textbf {\bibinfo {volume} {142}},\
  \bibinfo {pages} {391} (\bibinfo {year} {1981})}\BibitemShut {NoStop}%
\bibitem [{\citenamefont {Matthews}\ and\ \citenamefont
  {Stein}(1969)}]{matthews1969sensitivity}%
  \BibitemOpen
  \bibfield  {author} {\bibinfo {author} {\bibfnamefont {P.}~\bibnamefont
  {Matthews}}\ and\ \bibinfo {author} {\bibfnamefont {R.}~\bibnamefont
  {Stein}},\ }\href@noop {} {\bibfield  {journal} {\bibinfo  {journal} {Journal
  of Physiology}\ }\textbf {\bibinfo {volume} {200}},\ \bibinfo {pages} {723}
  (\bibinfo {year} {1969})}\BibitemShut {NoStop}%
\bibitem [{\citenamefont {Cleland}\ and\ \citenamefont
  {Enroth-Cugell}(1966)}]{cleland1966cat}%
  \BibitemOpen
  \bibfield  {author} {\bibinfo {author} {\bibfnamefont {B.}~\bibnamefont
  {Cleland}}\ and\ \bibinfo {author} {\bibfnamefont {C.}~\bibnamefont
  {Enroth-Cugell}},\ }\href@noop {} {\bibfield  {journal} {\bibinfo  {journal}
  {Acta Physiologica Scandinavica}\ }\textbf {\bibinfo {volume} {68}},\
  \bibinfo {pages} {365} (\bibinfo {year} {1966})}\BibitemShut {NoStop}%
\bibitem [{\citenamefont {Enroth-Cugell}\ and\ \citenamefont
  {Robson}(1966)}]{enroth1966contrast}%
  \BibitemOpen
  \bibfield  {author} {\bibinfo {author} {\bibfnamefont {C.}~\bibnamefont
  {Enroth-Cugell}}\ and\ \bibinfo {author} {\bibfnamefont {J.~G.}\ \bibnamefont
  {Robson}},\ }\href@noop {} {\bibfield  {journal} {\bibinfo  {journal}
  {Journal of Physiology}\ }\textbf {\bibinfo {volume} {187}},\ \bibinfo
  {pages} {517} (\bibinfo {year} {1966})}\BibitemShut {NoStop}%
\bibitem [{\citenamefont {Knight}(1972)}]{knight1972dynamics}%
  \BibitemOpen
  \bibfield  {author} {\bibinfo {author} {\bibfnamefont {B.~W.}\ \bibnamefont
  {Knight}},\ }\href@noop {} {\bibfield  {journal} {\bibinfo  {journal}
  {Journal of General Physiology}\ }\textbf {\bibinfo {volume} {59}},\ \bibinfo
  {pages} {734} (\bibinfo {year} {1972})}\BibitemShut {NoStop}%
\bibitem [{\citenamefont {Benda}\ and\ \citenamefont
  {Hennig}(2008)}]{benda2008spike}%
  \BibitemOpen
  \bibfield  {author} {\bibinfo {author} {\bibfnamefont {J.}~\bibnamefont
  {Benda}}\ and\ \bibinfo {author} {\bibfnamefont {R.~M.}\ \bibnamefont
  {Hennig}},\ }\href@noop {} {\bibfield  {journal} {\bibinfo  {journal}
  {Journal of Computational Neuroscience}\ }\textbf {\bibinfo {volume} {24}},\
  \bibinfo {pages} {113} (\bibinfo {year} {2008})}\BibitemShut {NoStop}%
\bibitem [{\citenamefont {Wong}(2020)}]{Wong2020.02.17.953448}%
  \BibitemOpen
  \bibfield  {author} {\bibinfo {author} {\bibfnamefont {W.}~\bibnamefont
  {Wong}},\ }\href {\doibase 10.1101/2020.02.17.953448} {\bibfield  {journal}
  {\bibinfo  {journal} {bioRxiv}\ } (\bibinfo {year} {2020}),\
  10.1101/2020.02.17.953448},\ \Eprint
  {http://arxiv.org/abs/https://www.biorxiv.org/content/early/2020/06/08/2020.02.17.953448.full.pdf}
  {https://www.biorxiv.org/content/early/2020/06/08/2020.02.17.953448.full.pdf}
  \BibitemShut {NoStop}%
\bibitem [{\citenamefont {Imaizumi}\ \emph {et~al.}(2018)\citenamefont
  {Imaizumi}, \citenamefont {Ruthazer}, \citenamefont {MacLean},\ and\
  \citenamefont {Lee}}]{imaizumi2018spontaneous}%
  \BibitemOpen
  \bibinfo {editor} {\bibfnamefont {K.}~\bibnamefont {Imaizumi}}, \bibinfo
  {editor} {\bibfnamefont {E.~S.}\ \bibnamefont {Ruthazer}}, \bibinfo {editor}
  {\bibfnamefont {J.~N.}\ \bibnamefont {MacLean}}, \ and\ \bibinfo {editor}
  {\bibfnamefont {C.~C.}\ \bibnamefont {Lee}},\ eds.,\ \href@noop {} {\emph
  {\bibinfo {title} {Spontaneous Activity in the Sensory System}}}\ (\bibinfo
  {publisher} {Frontiers Media, Laussane},\ \bibinfo {year} {2018})\BibitemShut
  {NoStop}%
\bibitem [{\citenamefont {von Holst}\ and\ \citenamefont
  {Mittelstaedt}(1950)}]{von1950reafferenzprinzip}%
  \BibitemOpen
  \bibfield  {author} {\bibinfo {author} {\bibfnamefont {E.}~\bibnamefont {von
  Holst}}\ and\ \bibinfo {author} {\bibfnamefont {H.}~\bibnamefont
  {Mittelstaedt}},\ }\href@noop {} {\bibfield  {journal} {\bibinfo  {journal}
  {Naturwissenschaften}\ }\textbf {\bibinfo {volume} {37}},\ \bibinfo {pages}
  {464} (\bibinfo {year} {1950})}\BibitemShut {NoStop}%
\bibitem [{\citenamefont {Kim}\ \emph {et~al.}(2010)\citenamefont {Kim},
  \citenamefont {Sripati},\ and\ \citenamefont {Bensmaia}}]{kim2010predicting}%
  \BibitemOpen
  \bibfield  {author} {\bibinfo {author} {\bibfnamefont {S.~S.}\ \bibnamefont
  {Kim}}, \bibinfo {author} {\bibfnamefont {A.~P.}\ \bibnamefont {Sripati}}, \
  and\ \bibinfo {author} {\bibfnamefont {S.~J.}\ \bibnamefont {Bensmaia}},\
  }\href@noop {} {\bibfield  {journal} {\bibinfo  {journal} {Journal of
  Neurophysiology}\ }\textbf {\bibinfo {volume} {104}},\ \bibinfo {pages}
  {1484} (\bibinfo {year} {2010})}\BibitemShut {NoStop}%
\end{thebibliography}%

\end{document}